\documentclass[runningheads]{llncs}
\usepackage[T1]{fontenc}
\usepackage{graphicx}
\usepackage{color}
\usepackage{hyperref}
\hypersetup{
    colorlinks=true,
    linkcolor=blue,
    filecolor=magenta,      
    urlcolor=cyan,
    citecolor=blue,
    breaklinks=true,
    pdftitle={Is Writing Prompts Really Making Art?},
    pdfpagemode=FullScreen,
}

\graphicspath{{Figures/}}
\begin{document}
\title{Is Writing Prompts Really Making Art?} 
%
%
\author{Jon McCormack\orcidID{0000-0001-6328-5064} \and
Camilo Cruz Gambardella\orcidID{0000-0002-8245-6778} \and
Nina Rajcic\orcidID{0000-0001-6501-5754} \and
Stephen James Krol\orcidID{0000-0002-9474-3838} \and
Maria Teresa Llano\orcidID{0000-0002-4898-1755} \and
Meng Yang\orcidID{0000-0002-3378-2606}}

\authorrunning{J. McCormack et al.}
%
%
\institute{SensiLab, Monash University, Caulfield East, Victoria 3145, Australia 
\email{\{Jon.McCormack, Camilo.Cruzgambardella, Nina.Rajcic, Stephen.Krol, Teresa.Llano, Meng.Yang\}@monash.edu}}
\maketitle              
\begin{abstract}
In recent years Generative Machine Learning systems have advanced significantly. A current wave of generative systems use text prompts to create complex imagery, video, even 3D datasets. The creators of these systems claim a revolution in bringing creativity and art to anyone who can type a prompt. In this position paper, we question the basis for these claims, dividing our analysis into three areas: the limitations of linguistic descriptions, implications of the dataset, and lastly, matters of materiality and embodiment. We conclude with an analysis of the creative possibilities enabled by prompt-based systems, asking if they can be considered a new artistic medium.

\end{abstract}
%
%
\section{Introduction}

We live in an age defined by technological innovation, while our world floods and burns with increasing veracity and severity. Over the last decade a seemingly endless wave of innovations in generative machine learning (ML) models have allowed the generation of photo-realistic images of non-existent people \cite{Karras}, coherent paragraphs of text \cite{vaswani2017attention}, conversion of text directly to runable computer code and, most recently from text descriptions to images \cite{https://doi.org/10.48550/arxiv.2204.06125}, video \cite{singer2022make}, and even 3D models. 

\begin{figure}
\begin{center}
 \includegraphics[width=0.8\textwidth]{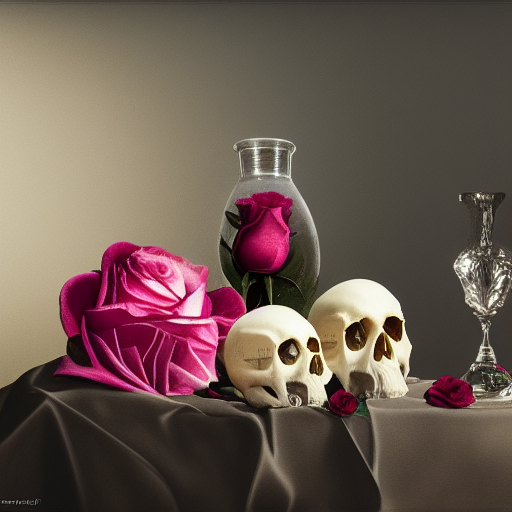}
\caption{Image generated by stable diffusion from the text prompt: \textit{``still life with human skulls of different sizes, a rose, the most beautiful image ever seen, trending on art station, hyperrealistic, 8k, studio lighting, shallow focus, unreal engine''}.} \label{fig:sd_example}   
\end{center}
\end{figure}

Systems such as DALL-E 2, MidJourney and Stable Diffusion allow the generation of detailed and complex imagery from short text descriptions. These Text-to-Image (TTI) systems allow anyone to write a brief English description and have the system respond with a series of images that depict the scene described in the text, typically within 5-30 seconds. An example is shown in Figure \ref{fig:sd_example}. Despite being quite recent and still an on-going development, these systems have become highly popular and an abundance of machine synthesised examples can be regularly found on social media, NFT sites, and other online platforms\footnote{MidJourney's interface is through the popular on-line messaging application \emph{Discord}, allowing anyone with permission to the appropriate channel access to both the prompts and generated images.}.

The obvious source of these systems' popularity is that they offer something entirely new: being able to generate an image just by describing it, without having to go to the trouble of learning a skill -- such as illustration, painting or photography -- to actually make it.  And importantly, the quality and complexity of the images generated is often comparable to what an experienced human illustrator or designer could produce. Moreover, TTI systems demonstrate a semantic interpretation of the input text and can convert those semantics so that (in some cases) they are more-or-less coherently represented in the generated images. This new-founded capability has inspired many useful image generation and manipulations possibilities, such as ``outpainting'' where a pre-existing image can have its edges extended with coherent and plausible content, or as an ``ideation generator'' where new versions of a set of input images are generated.

The makers of such tools pitch them as artistic, in a press release from LMU Munich (the University responsible for training Stable Diffusion), Björn Ommer, group leader of the Machine Vision \& Learning Group at LMU claimed that, ``the model removes a barrier to ordinary people expressing their creativity''\footnote{\url{https://www.lmu.de/en/newsroom/news-overview/news/revolutionizing-image-generation-by-ai-turning-text-into-images.html}}, and the first release of \emph{Stable Diffusion} claimed that it ``will empower billions of people to create stunning art within seconds''\footnote{\url{https://stability.ai/blog/stable-diffusion-announcement}}. 

As with many previous new technologies in the Arts, some artists have expressed  resistance to embracing them as legitimate artistic tools. This is unsurprising, initial efforts following the invention of the photographic camera resulted in much technical innovation and significant creative implications, but there was an initial reluctance to accepting photography as a legitimate art form. It took many decades for photography to gain acceptance within the Art Academy, and finally gain wide inclusion as a valid artistic medium. It is also important to note that using a camera does not necessarily make you an artist. Millions of people happily snap photographs with their smartphones every day, but the images captured are not typically presented as works of art. However, one important difference between TTI systems and other technologies adopted for artistic purposes is that TTI systems are specially pitched as being artistic tools, i.e.~ their primary intended application is to create ``art''.

In this paper we look at some of the implications and concerns expressed in this idea. We analyse the creative and artistic value of these new tools beyond the bold claims of press releases of those with a large commercial stake in the success of such models, looking at the implications of Text-to-Image creativity and the underlying ethical considerations of such systems. We unpack the drawbacks and the opportunities of TTI systems employed as artistic medium, offering an analysis of their potential for cultural contribution.

We divide our analysis of TTI systems into three key topics: the limitations of linguistic descriptions,  the implications of the dataset, and lastly, matters of materiality and embodiment in TTI systems. Following this analysis we conclude with a discussion around the artistic and creative opportunities enabled by treating TTI systems as a new artistic medium. 

\section{The Limitations of Linguistic Description}
\label{s:linguistics}
\begin{quote}
If you could say it in words, there would be no reason to paint
\flushright --- Attributed to Edward Hopper
\end{quote}

An obvious limitation of any system that attempts to interpret text into another medium (such as an image) is that one must be able to express the desired output linguistically (i.e.~\emph{in words}). We only need to look at art history to see that this is often very difficult or impossible, particularly -- but not exclusively -- for non-figurative images. Visual art has often claimed to be able to express the inexpressible \cite{Gombrich1995} -- to show through visual media what cannot be directly expressed in words. One only has to think of the works of Mark Rothko, Willem de Kooning, Clyfford Still or Bridget Riley as examples of this proposition. We note the important difference between an image expressing something that is beyond words and using words to describe a desired image (that may not yet exist).

In discussing what can be expressed linguistically, we need to differentiate that which cannot be expressed \emph{a priori} and which might be expressed \emph{a posteriori}. For example, phenomenological experiences cannot be fully expressed linguistically. Individual mental lexicons vary across human populations\footnote{For example a 20-year old native English speaker's vocabulary may range between approximately 27,000 to 51,000 lemmas.}. Furthermore, one may experiment with language in deriving appropriate text in a text to image system, learning from prior examples or growing their lexicon in response to working with such systems. 

\begin{figure}
\begin{tabular}{cc}
    \includegraphics[width=0.48\textwidth]{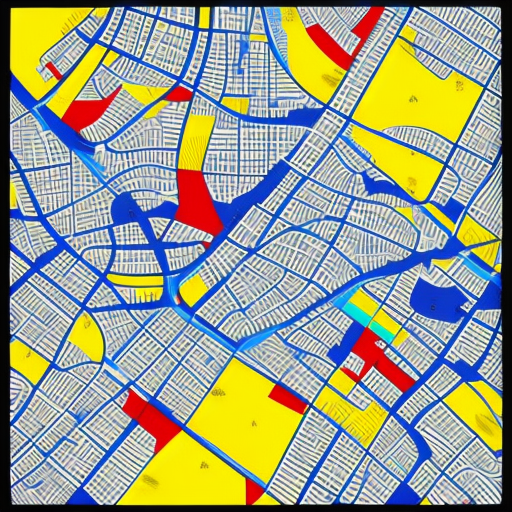} &
    \includegraphics[width=0.48\textwidth]{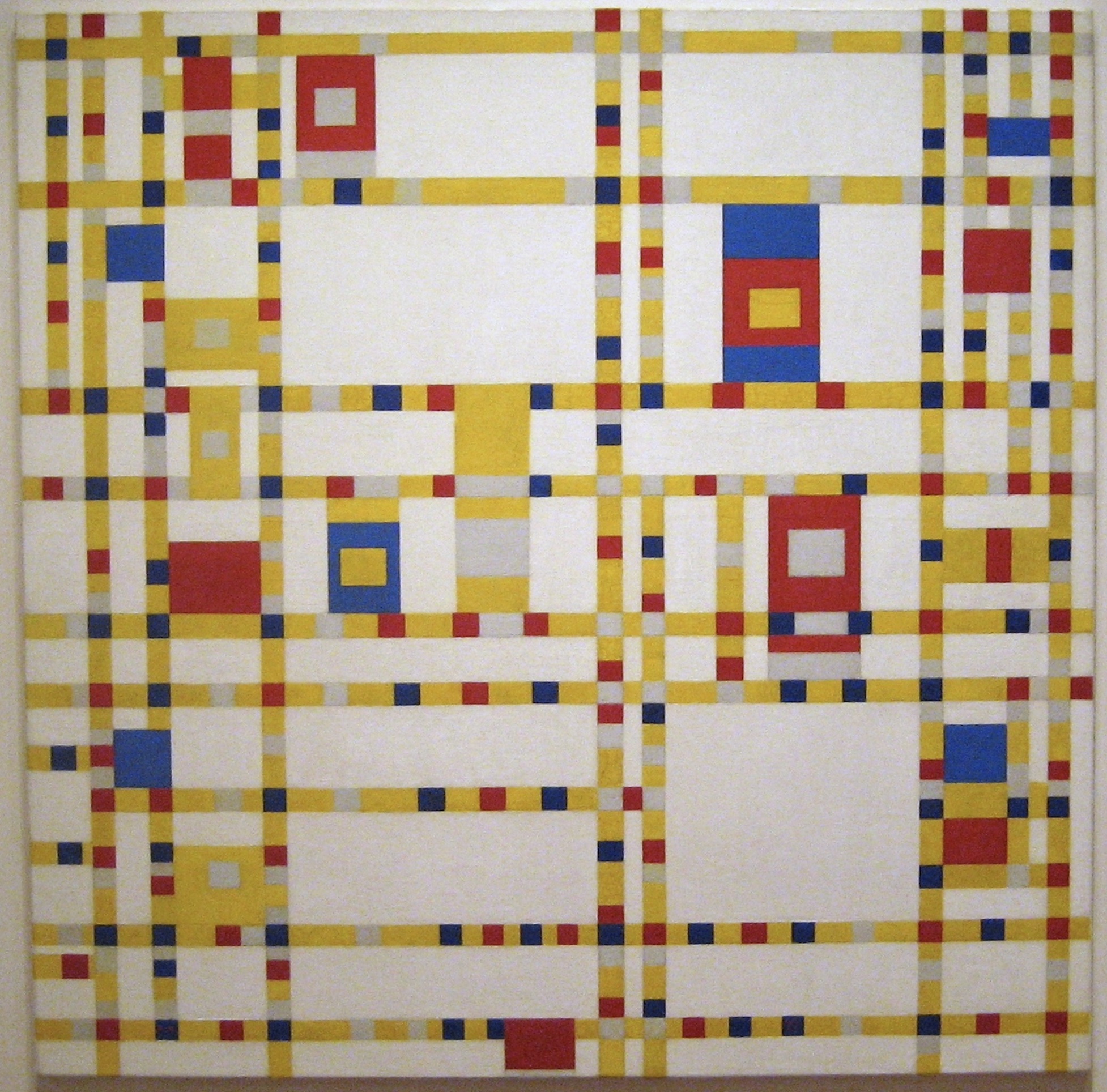} \\
     a & b 
\end{tabular}
\caption{Two maps: a) Broadway generated by Stable Diffusion from the text prompt: \textit{``An abstract map of Broadway in yellow, blue and red''}.. b) Broadway Boogie Woogie by Piet Mondrian (image credit: Wally Gobetz)} \label{fig:map}
\end{figure}

It could be argued that one may be able to create an image ``a-la-Rothko'' by providing an excruciatingly detailed set of physical instructions, an ``iconic'' form of representation that could be compared to the lists of commands used in \textit{turtle graphics}, where the meaning of the instructions has no relationship to the perceived meaning of what they produce. The appeal of TTI systems lies on their capability of translating symbolic representations -- textual descriptions -- into iconic ones -- images (see \cite{hisarciklilar2008symbolic}). However, the way this translation is implemented can be perceived as a limitation in two key ways:

 The ``understanding'' that TTI systems have of images is literal, as they are built upon millions of formal descriptions of visual material, where words or descriptive phrases (noun+adjective or verb+adverb pairings) capture the visual identity and characteristics of elements present in an image, their spatial relationships (e.g. ``a red flower in a vase'', ``a human skull on a table''), as well as some more general stylistic features (e.g. ``still life'', ``unreal engine''). Consequently the images produced by these systems can only reflect the content of a prompt literally. Any further meaning, intention or encoded information assigned to a specific textual construct, be it metaphorical or culturally charged, will be lost in translation. One example of this is how notoriously bad TTI systems are at producing diagrammatic or abstract images, where lines, annotations, dimensions, shading colour and texture are loaded with information beyond their mere graphic expression (Figure \ref{fig:map}).

Secondly, and especially under the paradigm of creative practice as a reflective process \cite{schon1984reflective}, in which designers and artists ``discover'' novel alternatives by following cues that emerge from in the work through their own creative actions. These come in the form of non-textual constructs -- diagrams, sketches, motifs -- that illuminate pathways to what things could become, by displaying new physical and/or perceptual order \cite{Alexander1964}. Explicitly verbalising something that has yet to be formalised -- i.e.~ex-ante description -- is akin to describing a colour that does not exist \cite{mcneill1972colour}.

Such limitations may explain why almost all the images produced by TTI systems are figurative and literal (e.g.~Figure \ref{fig:sd_example}). One might say that these systems are ``object-oriented'' in that they are trained predominantly on literal descriptions of images or using machine-trained image description technologies, such as CLIP, which identify objects in an image. Such systems have no understanding of metaphor, analogy or visual poetry, hence they cannot understand what is ``in'' an image in the way that a human artist or designer implicitly would\footnote{These limitations have been extensively documented in previous deep learning systems, a classic example being the painting by Margritte, \textit{La Trahison des Images}.}.

\subsection{Levels of Control}
\label{ss:levelsOfControl}

Another issue with such high-level text descriptions is the level of control and malleability one has over image making via language only. 
Painting, drawing, photography -- all human image-making techniques -- employ the body centrally in their production. This raises issues of embodiment, materiality, and material agency that traditionally come into play when making a visual artwork, something elaborated on in Section \ref{s:materiality}. In many aspects of art-making, modes of cognition such as verbalisation are not the primary means of art production, for example the paintings of Jackson Pollock reflect the interplay between body movement, paint and physics, which are all critically important.

\begin{figure}
\begin{tabular}{cc}
    \includegraphics[width=0.48\textwidth]{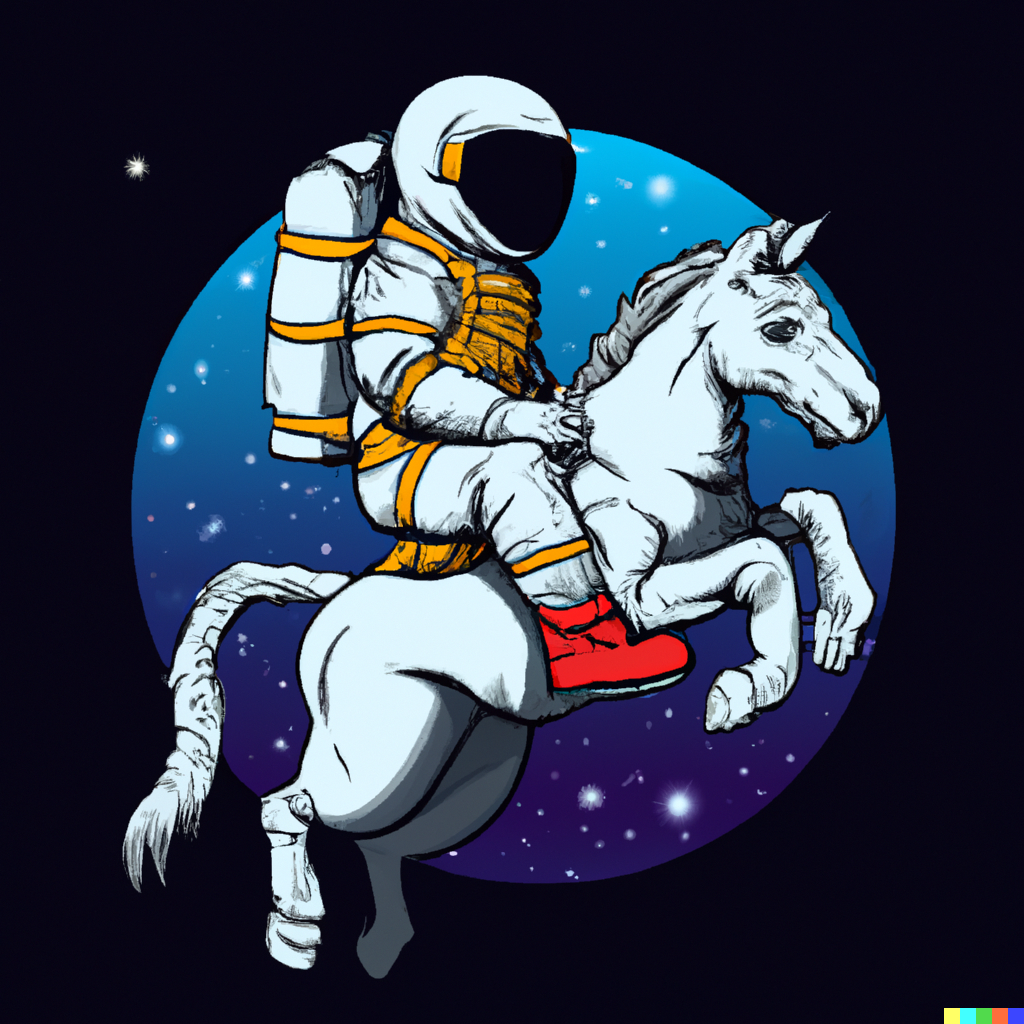} &
    \includegraphics[width=0.48\textwidth]{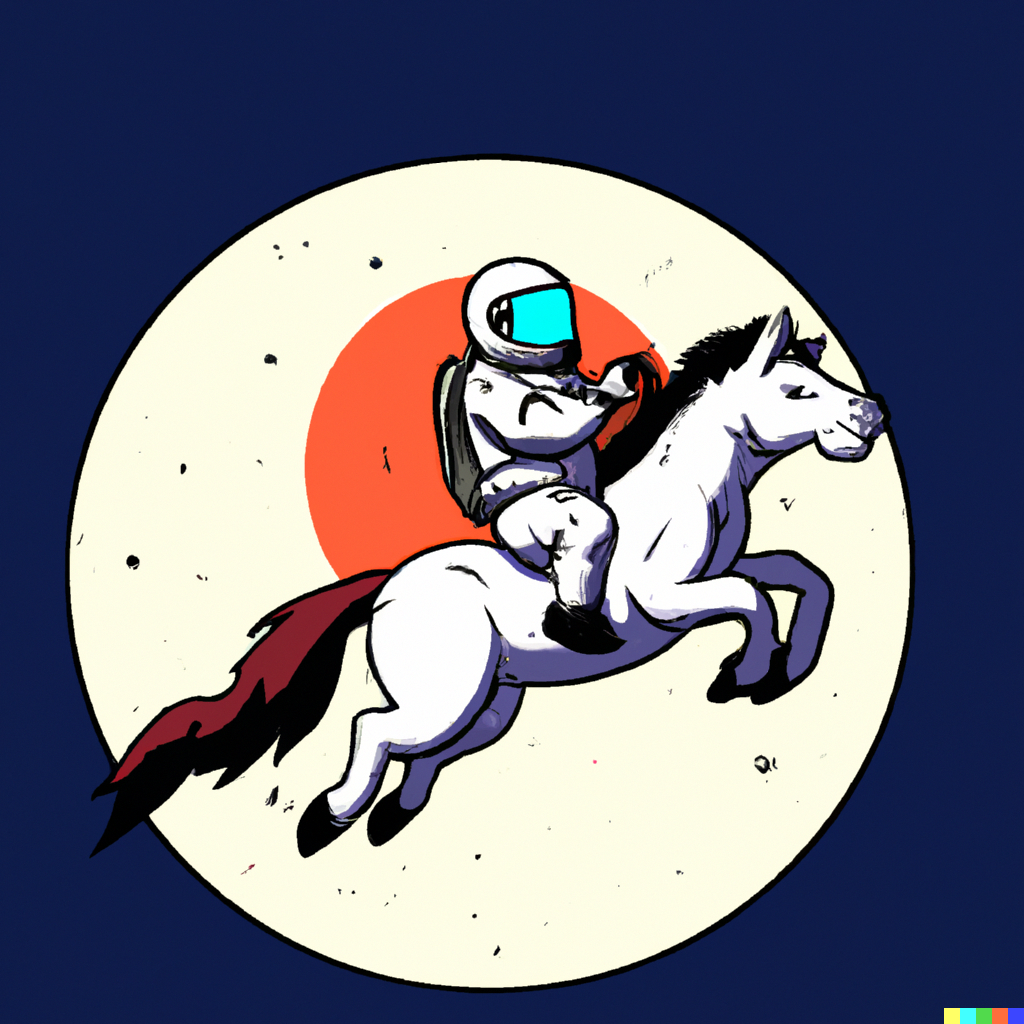}. \\
     a & b 
\end{tabular}
\caption{Two images generated using DALL-E 2: a) Text prompt: \textit{``an astronaut riding a horse''} b) Text prompt: \textit{``a horse riding an astronaut''}}
\label{fig:horseRiding}
\end{figure}

Even for those aspects of image production that TTI systems are able to control, their conceptual, relational and semantic understanding is not like that of a human \cite{Shanahan2022}. For example, prompting a TTI system for ``an astronaut riding a horse'' gives a literal representation of that description, but prompting for ``a horse riding an astronaut'' gives much the same imagery as seen in Figure \ref{fig:horseRiding}.

Issues of control are even more vexed when we consider the underlying training data and how systems establish statistical associations between words and images. Similar to Internet search engines, linguistic terms return the most common (or most populous in the dataset) interpretations of language-image associations. So if one asks a TTI system for a picture of a ``beautiful man'' the system returns an Internet stereotypical image of a man: white, heterosexual, youthful, professional, athletic. When used as part of prompts, basic artistic concepts such as ``beauty'' follow the statistical patterns expressed in the dataset, precluding cultural differences, homogenising representations and reinforcing biases.  

\subsection{Authorship}
\label{ss:attributions}

\begin{figure}
\begin{center}
\includegraphics[width=0.48\textwidth]{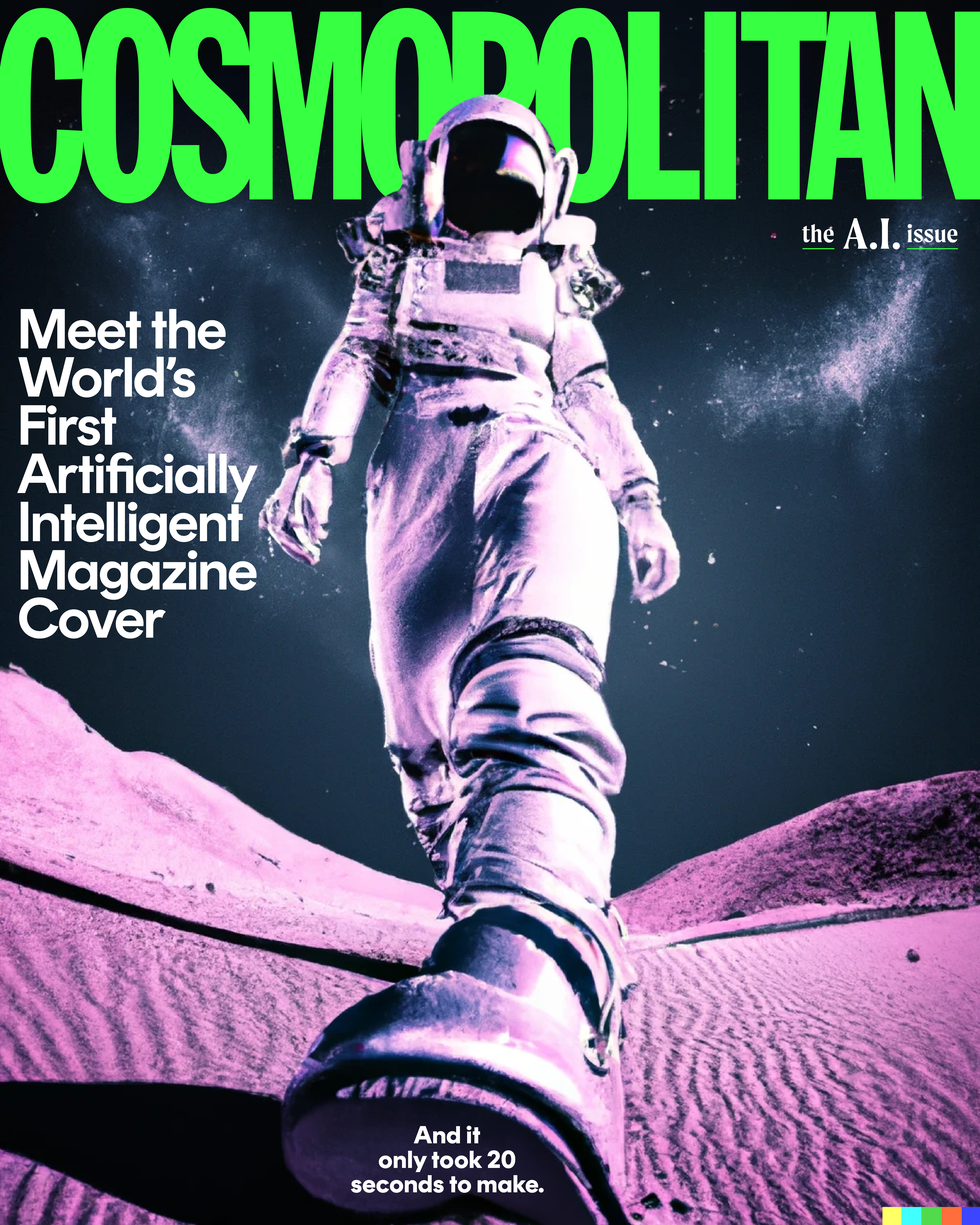}
\end{center}
\caption{Cosmopolitan cover created by digital artist Karen X. Cheng using DALL-E 2 and the text prompt \textit{``wide-angle shot from below of a female astronaut with an athletic feminine body walking with swagger toward camera on Mars in an infinite universe, synthwave digital art''}} 
\label{fig:cosmopolitanCover}
\end{figure}

A fundamental question in the debate about AI art has been the level of autonomy that computer systems have in making decisions that are essential for the creative process. In June 2022, Cosmopolitan magazine published an issue with what they claimed was ``The First Artificially Intelligent Magazine Cover'' (using DALL-E 2) which they further claimed ``only took 20 seconds to make'' (see Figure \ref{fig:cosmopolitanCover}). However, the article that describes the experiment in the magazine\footnote{\label{cosmopolitanArticle}\url{https://www.cosmopolitan.com/lifestyle/a40314356/dall-e-2-artificial-intelligence-cover/}} reveals that a digital artist and other members of the magazine's team had heavily intervened in the design and production of the cover, which actually took around 24 hours of work. 

This current hype around AI, particularly in the last couple of years, has shown an inclination to overstate the role of AI systems and understate the role of the human artist, exploiting the framing of ``autonomous AI artists'' (who are neither autonomous or artists)  for marketing and publicity reasons \cite{epstein:2020,notaro:2020,cetinic:2022}, and creating ethical issues around authorship.

\section{Data Implications}
\label{s:dataset}

\begin{quote}
    \textbf{parasite} \textit{(n)} an organism that lives in or on an organism of another species (its host) and benefits by deriving nutrients at the other's expense
\end{quote}

\subsection{AI as Parasite}
We argue that deep learning models trained on a large dataset of human created art -- such as TTI systems -- are \emph{parasitic} to human art and creativity, in that they derive and sustain themselves from human art, ultimately at the expense of current human art. In Biology, for a relationship between species to be parasitic, the parasite must benefit at the expense of the host. This is different from other relationships, such as symbiosis or mutualism, where both may benefit. The direct dependency of TTI models trained on human art datasets needs no further clarification, so to justify our claim of parasitism we need to demonstrate a negative effect on human art. Hence, if human art is diminished by TTI and similar systems then the relationship can be justified as parasitic.

As discussed in Section \ref{s:linguistics}, TTI systems reinforce the statistical norms and biases of their training sets through popular interpretations of language and images associated with them. Through this reinforcement (a kind of cultural colonisation), biases become normalised, less popular interpretations are more difficult to specify linguistically, so become marginalised, and eventually may be forgotten or lost. Additionally, the training data is \emph{backward-focused} in that it learns only from prior, existing imagery. Even more limiting is that this imagery must be accessible on the Internet. Of course, human art is not created \emph{ex nihilo}. Artists are influenced and inspired by those who came before them \cite{cynthia_close}. For example, Van Gogh used direct imitation in some of his work \cite{HomburgCornelia1996Tcto} and research has shown that viewing and imitating unfamiliar work is a powerful tool for art students to learn about art \cite{OkadaTakeshi2017IIaC}. However, unlike these TTI models, human artists draw from more than just previous art to inspire their own creations. A simple example is the depiction of nature in art. Van Gogh's work ``A Wheatfield, with Cypresses'' and Monet's ``The Water-Lily Pond'' not only draw from previous work, but are also inspired by the artist’s experience in those settings. This experience is more than just an image of a wheat field or a pond of lilies. It is a combination of emotion, feeling, mindset and other attributes that are difficult to convey linguistically.

Additionally, while TTI systems can generate ``new'' images (in the sense that the particular image has not existed before), they are all statistical amalgamations of pre-existing images, precluding the possibility of stylistic innovation, let alone any conceptual innovation (discussed further in Section \ref{s:materiality}). In the terminology of Boden \cite{Boden1991}, they support only combinatorial creativity (the combination of existing elements), not transformational creativity (where elements are transformed into something new).  In simple terms, TTI systems could not ``invent'' Cubism if they were only trained on data prior to 1908, yet this was possible for human artists.

If TTI systems become an increasingly significant part of cultural production\footnote{With major image creation platforms such as Shutterstock teaming with Stability AI it appears that AI synthesised images will increasingly become part of the cultural vernacular.} it is possible that human art and creativity will be diminished for the reasons given above. Moreover, as such systems replace human illustrators the pool of human talent able to competently illustrate may diminish as paying a human illustrator is less viable commercially than using an AI.

\subsection{Parasitic Meaning}
\label{ss:parasiticMeaning}

This idea that generative computer algorithms that mimic human art are ``parasitic'' isn't new. The philosopher Anthony O'Hear formed this view almost 30 years ago \cite{Ohear1995}. O'Hear considered computer programs that could generate art similar to human art, first emphasising the need to understand that appreciation of visual art was more than just the consumption of visual images. An important criteria of something being considered ``art'' meant that there was a direct connection between art as a product and its ``history or mode of production'' \cite{Ohear1995}. Hence, art was \emph{more} than just pleasant visual appearances or surface aesthetics, it relied on human \emph{intention}: the intention to engender a particular experience, to communicate an idea, or elicit a particular emotional response. O'Hear conceded that not \emph{every} work of art necessarily achieved this; there may be ambiguity or some other failure where the intention wasn't realised for the work's audience. Nevertheless, in O'Hear's view, this intention and authenticity were necessary conditions for something to be considered as ``art'', as opposed to the more general class of visual images that saturate contemporary life.

O'Hear specifically considers the possibility of computer-generated works of art that ``drew on repertoires derived from existing works, but which were not copies of any complete works''. He argues that any meaning(s) attributed to them come from those that originated through human thought. Hence they ``would be parasitically meaningful, deriving their meanings from the techniques and conventions which human artists had developed in their works.'' \cite{Ohear1995}.

We are not proposing that TTI systems are unable to communicate human intention, rather that much (or all) of that intention is derived from preexisting human art, hence any meaning inferred is parasitic, in O'Hear's sense of the term. The author of the prompt that generated an image may also have some intention (the intention to make a ``beautiful'' image, to combine or juxtapose unlikely elements, to create something that mimics an existing human style, even to convey a particular feeling or mood), but this intention can only ever be peripherally enacted in the image itself, since it is the TTI system, not the prompt author, that is actually making the image.

\subsection{Data Laundering}

One of the key factors that contributes to the capability of TTI models is their access to massive datasets used for training and validation. Achieving the visual quality and diversity that they are capable of reproducing requires a very large corpus of human-created imagery, which is typically scraped from the Internet. In a practice that has been dubbed ``data laundering'', scraped datasets -- which include large amounts of copyrighted media -- rely on special exemptions for ``academic use'' to avoid any legal barriers preventing their use, or for copyright owners to claim against \cite{data_laundry}. For example, Stability AI (the ``creators'' of Stable Diffusion) funded the Machine Vision \& Learning research group at the Ludwig Maximilian University of Munich to undertake the model training and a small nonprofit organisation, LAION, to create the training dataset\footnote{https://laion.ai/projects/} of approximately 5.85 billion images, many of which are copyrighted, and in general appropriated for this purpose without the image creator's direct permission.

Concern has been raised by artists about the ethical and moral implications of their work being used in such systems. These concerns include the appropriation of an individual artist's ``style'', mimicry, and even the replacement of a human artist or illustrator. Furthermore, there is no easy way to be excluded or removed from such datasets. Such is the concern raised by TTI systems that websites such as \url{https://haveibeentrained.com/} have emerged, allowing anyone to search to see if their work has been used as part of the training data. A companion project allows people to opt-in or opt-out of their data being included. At the time of writing (November 2022) around 60\% of respondents chose to opt-out of being included in any training data. 

The use of copyrighted images in datasets creates another issue, raising the question that whether training models on copyrighted data should be considered plagiarism. Being able to easily generate an image in an artist's style without paying for that artist to create it (or paying any royalties or licensing fees), allows users of such technology to bypass the traditional economic, legal and moral frameworks that have supported artists traditionally. Generating copyright-free images immediately for commercial use without the cost or time involved in securing copyright from a human artist may become an attractive proposition, raising the interesting legal question of who would be the defendant in any copyright infringement case brought about by this scenario. 

\section{Materiality and Embodiment}
\label{s:materiality}

\begin{quote}
     AI can't handle concepts: collapsing moments in time, memory, thoughts, emotions -- all of that is a real human skill, that makes a piece of art rather than something that visually looks pretty.
     \flushright --- Anna Ridler
\end{quote}

The vast majority of visual art currently made by humans is made \emph{physically}, that is it involves embodied physicality, situated awareness, and the interplay of material agencies. As expanded upon in Section \ref{s:linguistics}, TTI systems rely solely on language for the production of digital images, excluding any possibility of interaction that precedes representation. This limited interface excludes the phenomenological and embodied modalities that typically comprise creative practice, presenting a number of implications for TTI systems as artistic medium.

Central to human creativity is the cultivation of an artistic practice. The honing and refinement of an artistic craft contributes not only to the quality of the work, but also embeds tacit knowledge within the wider community, shaping the direction of future work created in a given medium. In the case of TTI systems, the artistic ``practice'' involves the iterative prompting and re-prompting of the TTI model, often touted by AI artists as veritable skill which has been generously dubbed ``prompt-engineering''\footnote{This ``skill'' maybe short-lived: models already exist to translate between descriptive text and ``prompt text'' \cite{Microsoft2023}.}. Some prompt engineers are unwilling to reveal their specific prompts for fear that others may easily generate the same results, thereby diminishing the ``originalilty'' of the imagery they have been able to find. Indeed, the refinement and elaboration of the text prompt does lead to ``better'' (a subjective measure) images, or at least grants one more control over the resulting image.

Prompt-engineering falls short as an artistic practice for a number of reasons. Firstly, although the skill is developed through countless hours spent engaging with the medium, it is not learned in any tacit, embodied, or individualised way. Prompts can be directly shared, copied, stolen, and in an instant the entire body of knowledge and invested time is transferred. Prompt-engineering, if shared, does indeed contribute knowledge towards the broader community, yet holds little personal value as the art object. 

This is not to say that writing prompts could \emph{never} be considered an artistic practice. Any medium has the potential to be used artistically, and as software art and creative coding demonstrate, the authoring of a process may embody significant artistic qualities. However for prompt writing, there appears to be a (low) ceiling in skill development, which is unusual for an artistic medium, and is likely due to the highly limited possibilities for interaction provided by current systems. 

Finally, and perhaps most importantly, prompt writing is entirely confined to a screen. This does not preclude its potential as an artistic medium, but rather brings into question whether this is the kind of artistic practice we as a community wish to foster, particularly if that comes at the expense of others. TTI models themselves are not only divorced from the physical world but are in turn emancipating artists from it too. This is further analysed in the following sections.

\subsection{Embodiment in AI}
There is a vast amount of research on understanding and defining artistic practice within artificial environments. A significant component concerns discerning the meaning of ``embodiment'' for AI systems, something that we reflect upon in this section.

Although the prevalent view of embodiment has established that it cannot exist in an ``abstract algorithm'' \cite{pfeifer:2001} and has emphasised the importance of physical grounding \cite{brooks:1990}, others have put forward the idea that a physical body is not a fundamental requirement, and argue that embodiment is given by an agent {\em being situated} in an environment. Hence a {\em virtual} environment that provides a (high-fidelity) simulation of a physical world can enable embodiment, provided the agents are {\em structurally coupled} with that environment (i.e. the agents and their actions are bounded to the structure of the environment) \cite{franklin:1997,guckelsberger:2021}. Systems like DALL-E, Midjourney and Stable Diffusion, and in general TTI systems, do not exist in a virtual environment that enables any kind of agent-environment interaction. They learn statistical patterns from the data and generate new images altering these patterns, but they do not independently grow or evolve over time, they do not employ any mechanisms of change. 

Cognitive scientists have also expanded the notion of embodiment to not only refer to the physical body but also to other aspects of an agent's interaction with the environment. In a classification of different notions of embodiment, Ziemke \cite{ziemke:2003s} highlights the notion of {\em historical embodiment}, which refers to the ``result or \textit{reflection} of a history of agent-environment interactions''. The main premise behind this, is that a system's state is shaped by both its present interactions with the environment and its experiences (or history) from the past. This definition has been highlighted by some as one that ``does not exclude domains other than the physical domain'' \cite{riegler:2002}. By this definition, we might say that TTI systems could be considered embodied as they are trained on digitized versions of historic pieces of artwork; however, a lot (if not most) of the information about the artist-environment interactions and the physical media itself, which were intrinsic in the creation of these artworks, are not part of the training data. TTI systems focus on the final artefact (the pixelated, visual representation), with little or no information about the cultural and social environment of the time, the materials, creative process, etc. 

When creating artworks, the boundary between the environment and artist is transformed, allowing both to experience each other, to move within each other's features and constraints, and adjust to them. TTI systems reinforce that boundary, significantly limiting the interaction between the artist and the system, as well as any experience of the system with the environment. It can be said that TTI systems’ ``experience'' of the world is limited to their training process, decoupling them from the environment and setting a hard boundary with their users during the generation process. As argued in \cite{cetinic:2022}, this makes it ``difficult to estimate if the value of a particular AI artwork should depend on the technological complexity and innovation involved in its production, or only on the final visual manifestation and contextual novelty''.

\subsection{Exposing Process}
\label{ss:exposingProcess}

The black-box nature of TTI generators (and Deep Learning models more generally) leads to a privileging of output over \textit{process}. The algorithm does not reveal it's process in producing output, nor is the process contained within the resulting image. Each new generated image is entirely independent of the images that came before it. In this way, we see a privileging of the visual image over its means of production. When employed in the conventional fashion, TTI systems generate visual imagery that is entirely divorced from process, artistic craft, or conceptual foundation; they fetishise surface aesthetics while often lacking a narrative component.

All of this is not to say that TTI and generative AI systems lack a materiality entirely. Of course, these systems are ultimately comprised of physical materials. The material entanglements of TTI systems stretch back to include echoes of the dataset (as elucidated in Section \ref{s:dataset}), and forward in their generative capacities and influence on visual and computing culture. 

\subsection{Material Agency}

In his seminal paper, Malafouris puts forth an argument for distributed and emergent agency with the example of a potter moulding clay \cite{malafouris2008potter}. He argues that the potter not only exerts an agency onto the clay, but that through the clay's resistance to being moulded it also exerts agency onto the potter. Hence, agency is not a pre-existing quality but emerges from within an interaction.
This notion of material agency has commonly been interpreted in relation to artistic media and practice. The qualities and constraints of a given medium are central to the artistic process; to create is to operate at the boundary -- as Malafouris calls it, the Brain Artefact Interface (BAI).
The information at the heart of digital systems may be incorrectly viewed as immaterial, yet ``just as molecular materials act as resistance and come to transform action upon material objects, so digital materiality comes to enable and transform creative practices upon computers'' \cite{poulsgaard2020understanding}.

In considering TTI models as a creative medium, we can see that it is precisely the resistance of the medium to being \emph{moulded} that leads to it's true creative potential. In a large majority of work created with TTI models we observe an attempt to conceal the qualities of the medium -- to render it invisible -- often by imitating more traditional materials like paint, photography, or with aesthetic conventions such as ``cinematic lighting''. According to a media-specific analysis popular to modernist thought \cite{carroll1985specificity},  it is through the abandoning of imitation and illusion that the full artistic potential of a given medium can be achieved. For example, the abstract art movement saw surrendering to the 2D plane, an escape from emulating the conditions of 3D media. In the same way we argue that TTI systems can become creatively interesting once we embrace their unique properties, rather than work against them or try and mimic popular aesthetic conventions.

\begin{figure}[h]
\begin{center}
\includegraphics[width=0.6\textwidth]{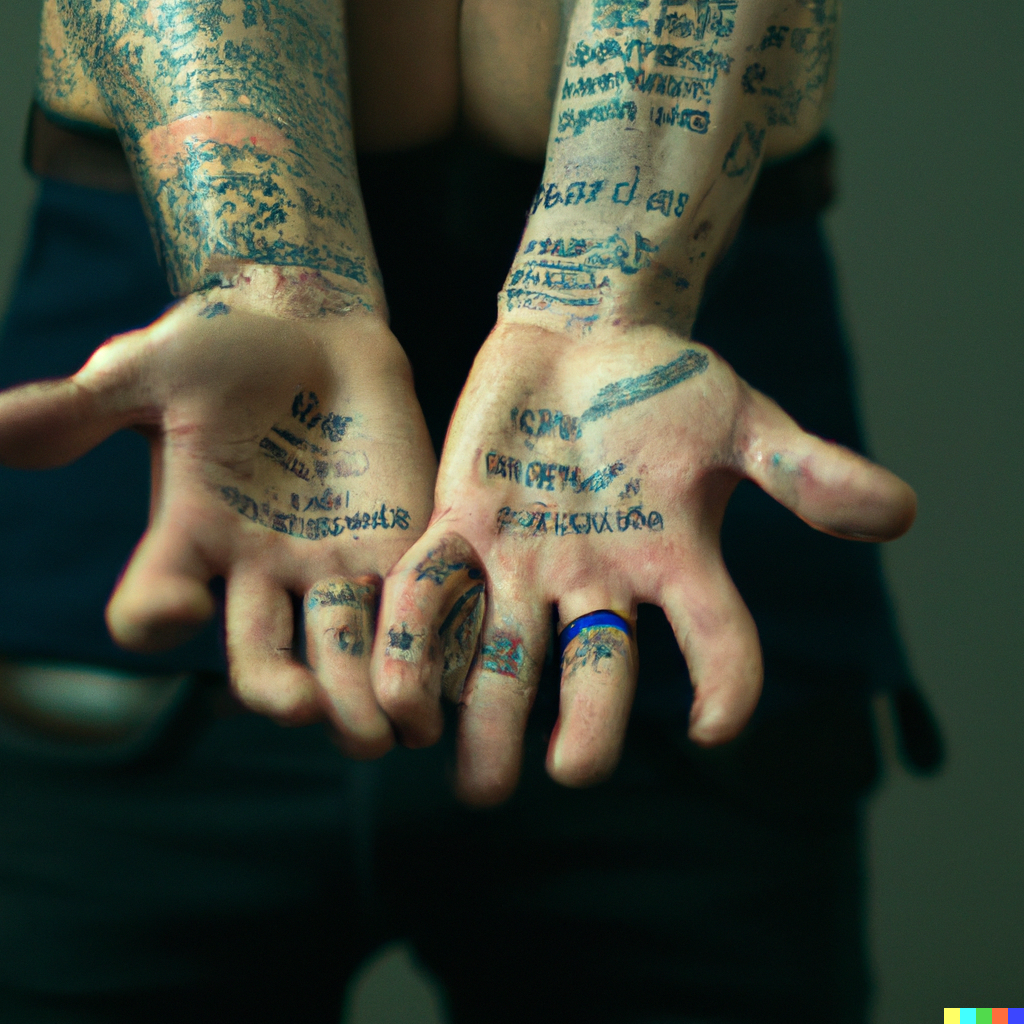}
\end{center}
\caption{Image generated by DALLE-2. Prompt: \textit{a man covered in tattoos of English words, long hair, rings on fingers, cinematic lighting, 8k,}} 
\label{fig:tattoos}
\end{figure}

\subsection{A New Medium?}

\begin{quote}
    Whatever you find weird, ugly, uncomfortable and nasty about a new medium will surely become its signature\ldots It's the sound of failure: so much modern art is the sound of things going out of control, of a medium pushing to its limits and breaking apart.
    \flushright -- Brian Eno
\end{quote}

While we argue that simply providing a prompt to a TTI system and taking its output does not constitute a rich artistic practice, we appreciate the possibility and potential of deep learning systems as an artistic medium. Duchamp's ``Fountain'' demonstrates how easily our understanding of what art is can be disrupted.

Rodolfo Ocampo, an AI artist, recently had some of their work showcased at a gallery in Sydney, Australia \cite{panagopoulos}. This work involved a portrait, generated by DALLE-2, of a person that does not exist. The motivation of the work was to see if people could build a connection with a non-existent subject. Generating people that do not exist is not a new concept \cite{Karras} (indeed it has been practised in painting and photography for centuries) and there may still be debate on whether this work constitutes art. However, the use of DALLE-2 in this setting is more nuanced than simple image generation as it is employed within a broader artistic concept -- building connections with non-existent AI generated humans.

It is common for AI artists to try and hide or fix errors generated by TTI systems. But exploiting these limitations might lead to more interesting artistic possibilities. An example of these errors can be seen in figure \ref{fig:tattoos}. The image portrays a subject covered in tattoos and the prompt specifies that the tattoos are English words. However, while the tattoos do have word-like qualities to them, none of them are in English (or in any human language). This demonstrates a common deficiency with TTI systems -- they are hopeless at generating meaningful text. Other ``errors'' in figure \ref{fig:tattoos} include the misshapen hands (a common issue for TTI systems) and the subject facing away from us even though the hands suggest he is facing towards us. These ``errors'' highlight interesting properties about the medium, revealing its limitations and raising questions about the underlying technology.


\section{Conclusion}

Technology has repeatedly transformed the arts and culture. Throughout modernity, many lamented the automation of artistic processes and practices due to technology (``Why bother making a painting when a camera will give you an image instantly''). As we sacrifice direct control and defer human intelligence to machines, we promote a shift that downgrades technique but promotes new sensibilities. People still bother making paintings because painting offers things that photography cannot. 

The title of our paper asked the question: ``Is writing prompts really making art?''. As we have discussed, prompt writing raises many important concerns: conceptual, ethical, legal and moral. These concerns should make us wary of blindly accepting generative machine learning systems as a positive thing for human creativity and art.

If you gave a designer or illustrator a verbal brief (and paid them) to create a specific illustration for you, you would not credit yourself as the creator or artist in such a scenario. The creator would be the artist who drew the work. With TTI systems there is a broader question of authorship, because the data used for training directly involves literally billions of examples of human authorship.

Hence, at this point it is difficult to see prompt writing as a significant art practice. However, as artists have capably demonstrated, anything can become a medium for art and artistic practice, so we expect to see artists making use of these new technologies for new creative purposes. As we have outlined, this will most likely be by subverting their intended use, embodying the agency and interaction for these systems or just embracing their flaws and quirks to expose more of the process of deep learning and machine creativity. Moreover, the form and modalities of current TTI systems will likely change rapidly and unpredictably through the widespread technological, engineering, political and cultural advances driving these systems.

\subsubsection{Acknowledgements} 
This research was supported by an Australian Research Council grant DP220101223.

%
%
%
\bibliographystyle{splncs04}
\bibliography{McCormack}

\end{document}